\documentclass[%
 reprint,
 amsmath,amssymb,
 aps,
]{revtex4-2}

\usepackage{graphicx}
\usepackage{dcolumn}
\usepackage{bm}
\usepackage{comment}
\usepackage{xcolor}

\begin{document}

\preprint{APS/123-QED}

\title{From wicking to anti-wicking: A universal framework for capillary dynamics}
\author{Aniruddha Saha}
\email{as2742@cornell.edu}
 \author{Sadaf Sobhani}
\affiliation{Department of Mechanical and Aerospace Engineering, Cornell University, Ithaca, NY 14850, USA
}%


\begin{abstract}
The dynamics of capillary rise under different geometric and fluid conditions have the common signatures of rapid rise followed by an equilibrium state that describe the underlying competing forces. We present a new interpretation of capillary dynamics using a linear damped system where modulation of damping and forcing characteristics are achieved using axisymmetric channels with sinusoidal variation in radius. The complete axisymmetric design space ranging from hydrophilic channels that enable spontaneous imbibition to hydrophobic channels, that required external pressure mechanisms is modeled and the force dynamics is split into simultaneous damping and forcing characteristics. We introduce the product of damping and forcing terms as the new parameter that effectively characterizes rise dynamics across various geometric and flow conditions, encompassing both flow-enhancing and flow-inhibiting scenarios. The monotonic nature of this parameter enables the development of a stochastic optimization method that can determine optimal channel geometries for controlled capillary rise.

\end{abstract}

\maketitle
 The wicking dynamics of fluid through narrow spaces is affected by the physical interactions with the bounding surfaces. This phenomenon is observed in natural as well as architected systems where the fluid moves through confinements in an attempt to minimize the free energy of the system. The representation of flow through porous media has been simplified as a capillary imbibition model and the Lucas-Washburn equation describes the vertical rise in such capillary channels \cite{lucas, lucas-washburn}. Capillary phenomenon is observed in a wide range of scenarios from biological applications \cite{tani2014capillary} to three-dimensional porous media \cite{henryk} where channels often feature irregular cross-sections and varying geometries. In non-cylindrical capillaries, interfacial flow relies on the local contact angle, when the presence of localized curvature in such channels introduces additional complexities to the flow regime \cite{tsori2006discontinuous,binjan}. Geometric parameters have been utilized as ``passive'' mechanisms to expedite liquid rise through capillaries \cite{baptiste, dudukovic2021cellular}. Similarly, the slower rise of liquid in sinusoidal channels have been observed using numerical simulations \cite{lbz}. While viscous dissipation typically slows liquid rise, channel geometry can be strategically manipulated to control flow behavior \cite{erickson2002numerical}. Constricted sections have been demonstrated to act as global flow resistors, while enlarged sections create localized resistance, making them valuable for regulating capillary flow velocity in architected biological applications \cite{berthier2016capillary}. Additionally, flexible walls in 2-D microfluidic channels have been shown to accelerate imbibition as the passing meniscus pulls on the wall and increases curvature \cite{anoop2015capillary}. Tapering of channels have been observed to improve flow rates up to a critical tapering angle, beyond which liquid cannot reach the channel tip \cite{baptiste} - a geometry naturally observed in plant xylems that transport liquid over large distances \cite{anfodillo2006convergent}.
 Despite the diversity in capillary geometries, Newtonian liquids generally exhibit characteristic power-law behavior ($h\propto t^\beta$) during wicking. At shorter time scales, the flow progresses through two distinct regimes: an initial inertial phase where $h \propto t$, followed by viscous diffusion where $h \propto \sqrt{t}$. However, at longer time scales, it has been shown that capillary geometry can significantly influence these power-law dynamics, leading to deviations from classical scaling exponents \cite{reyssat2008imbibition}. Interest in channel geometries that promote wicking is driven by the aim of enhancing passive fluid transport~\cite{figliuzzi2013rise}. Prior studies optimizing the rise times in capillaries have focused on maximizing the volume of liquid passing in a given time or minimizing the rise times over a defined distance \cite{baptiste}. Such methods impose these parameters on the desired cost function but the commonality in the underlying physics is not explored. 
\par In this letter, we seek to uncover the universal framework behind such modulation, offering a new perspective on flow control that encompasses both wicking enhancement and retardation that represents the anti-wicking tendency of the system. We present a universal description of capillary rise dynamics under a broad design space and have identified a parameter that accounts for both the flow and geometric contributions, enabling optimization of channels that can enhance or inhibit wicking. 
 \par We model a capillary with a cross section radius of $R(x)$, where $x$ represents the axial position along the length of the capillary, and length $l$. The liquid front is at height $h$, with air in the remainder of the channel. The density and viscosity of the liquid and gas are denoted by $\rho$, $\mu$ and $\rho_g$, $\mu_g$respectively. The contact areas between the solid-liquid pair and the liquid-gas pair contributes to the total free surface energy, which are then balanced by the changes in gravitational potential energy and viscous dissipation. To describe axisymmetric capillary flows, we write the rate of change of the free surface energy as:
 \begin{eqnarray} 
\delta \dot{E} = 2\pi\sigma\left[R \frac{dR}{dt}-\cos \theta\frac{d}{dt}\int_{0}^{h} R \sqrt{1+\left(\frac{dR}{dx}\right)^2}dx\right],~~~
\label{eq:freesu}
\end{eqnarray} where the local contact angle is $\theta$ and surface tension $\sigma$. The area integral is evaluated over the contact surface between the wall and the liquid, as well as the cross-sectional area at the meniscus, with local curvature effects due to the contact angle being neglected. The lubrication approximation is valid with the approximation of a parabolic velocity profile at the liquid front for channels exhibiting gradual changes in cross-section, where $dR/dx \ll 1$. The change in total energy is reflected through the changes in gravitational potential energy and the viscous work done by the fluids as they move through the channel. We can write the total change in energy of the fluids as:
\begin{eqnarray} 
    \delta \dot{E} &&= \frac{(\rho-\rho_g)g\delta \dot{E}}{\frac{2\sigma}{R}\cos{\left(\theta+\tan^{-1}\left(\frac{dR}{dh}\right)\right)}}h \nonumber\\ &&- 8 \pi R(h)^4\left[\int_{0}^h \frac{\mu}{R(x)^4}dx +\int_{h}^l \frac{\mu_g}{R(x)^4}dx\right]\dot{h}^2.
    \label{eq:rhs}
\end{eqnarray}
The final governing equation of capillary rise is derived by equating Eq.~(\ref{eq:freesu}) and ~(\ref{eq:rhs}). We then  non-dimensionalize the resulting equation. To generalize the description to include channels resisting spontaneous imbibition, we consider a typical contact angle that can be either hydrophilic ($\theta < \pi/2$) or hydrophobic ($\theta > \pi/2$). An external pressure is introduced to drive the flow under hydrophobic conditions, defined as a linear time-varying pressure ramp, given by $P=P_A+P_Bt$. The system is non-dimensionalized using the following parameters: $\bar{h}=h/l,~\bar{R}=R/R_0,~\bar{P}=P/(\sigma/R_0),~\bar{t}=t/(l^2\mu/\sigma R_0)$. The non-dimensional viscosity $\bar{\mu}$, which combines the viscous response from both liquid and gas phases, is defined as:
\begin{eqnarray} 
\bar{\mu} = \int_{0}^{\bar{h}} \frac{1}{\bar{R}(\bar{x})^4}d\bar{x} +\frac{\mu_g}{\mu}\int_{\bar{h}}^1 \frac{1}{\bar{R}(x)^4}d\bar{x}.
\label{eq:visc}
\end{eqnarray}
The continuously changing channel radius leads to a local angle $\alpha (h)=\tan^{-1}(\frac{dR}{dh})$. This angle quantifies the slope of the channel wall at any given height, reflecting the gradual variation in the channel's cross-sectional shape. We use the simplifications described above to represent Eqs.~(\ref{eq:freesu}),~(\ref{eq:rhs}) and ~(\ref{eq:visc}) in a dimensionless compact form as:
\begin{eqnarray} 
\frac{d \bar{h}}{d\bar{t}} =-D(\bar{h})\bar{h} + F(\bar{h}).
\label{eq:damped}
\end{eqnarray}
The above equation closely represents a linearly damped system without inertial contributions. The two key parameters: the damping coefficient $D$ causes the system to decay and forcing function $F$ drives the system are defined as:
\begin{eqnarray} 
D &=\frac{1}{8\bar{\mu}}\text{Bo} \left(\frac{l}{\bar{R}}\right) \left(1 - \frac{\rho_g}{\rho}\right) \frac{\cos(\theta) - \sin(\alpha)}{\cos(\alpha)\cos(\theta+\alpha)},\nonumber \\ 
F &=\frac{1}{8\bar{\mu}}\left(\bar{P} + 2 \frac{\cos(\theta) - \sin(\alpha)}{\bar{R}\cos(\alpha)}\right),
\label{eq:DF}
\end{eqnarray}

where the Bond number $\mathrm{Bo}=\frac{(\rho-\rho_g)gR_0^2}{\sigma}$ captures the balance between gravitational and surface tension forces. These parameters highlight the universal nature of capillary rise dynamics: a rapid initial rise followed by viscously dominated equilibration. 

\begin{figure}[t]
\includegraphics[width=0.48\textwidth]{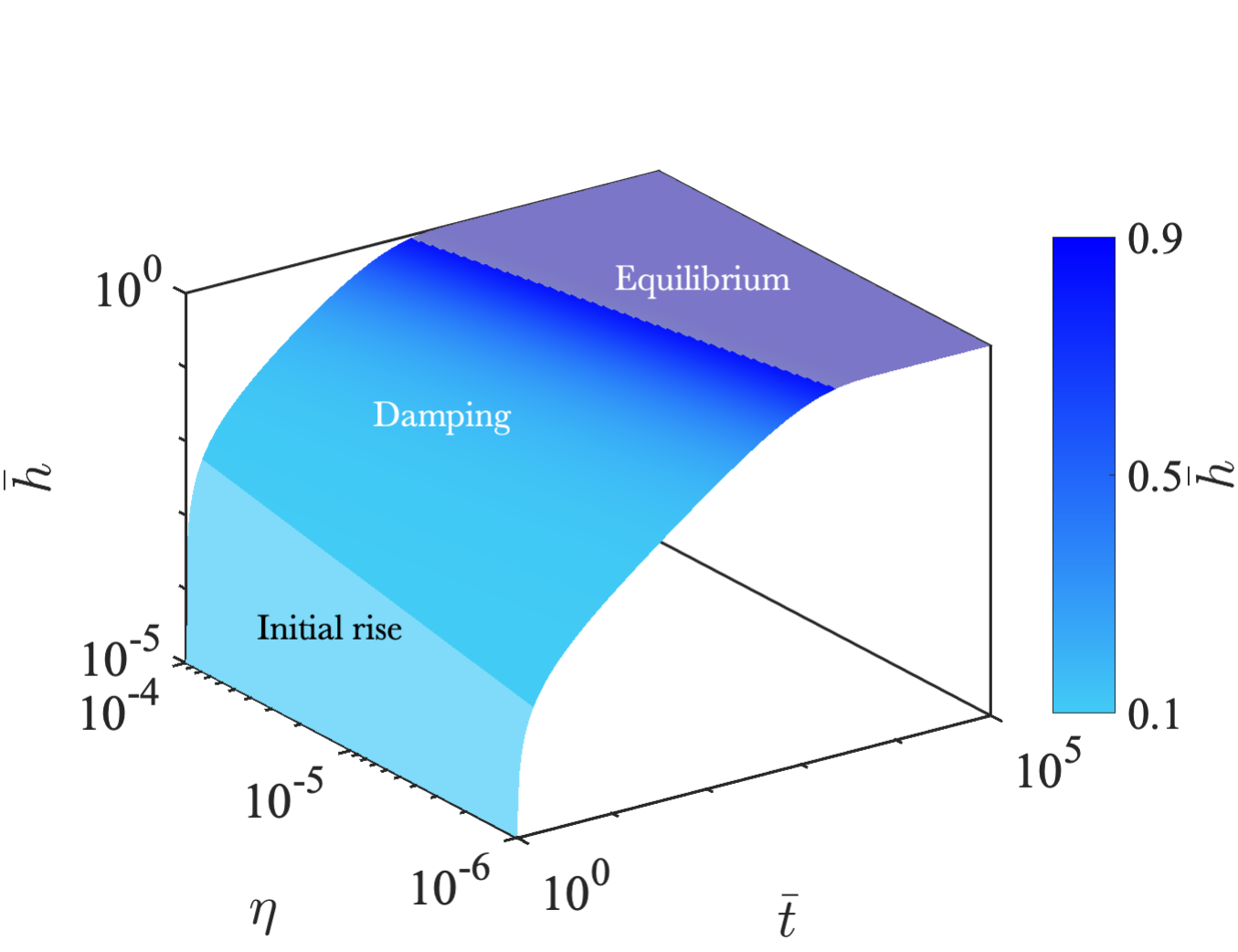} 
\caption{Solution of $\dot{\bar{h}}=-\mathcal{D}\bar{h}+\mathcal{F}$ shown for different values of the parameter $\eta = \mathcal{D}\cdot \mathcal{F}$. The regions of initial rise and equilibrium are shown at constant $\bar{t}$ and $\bar{h}$ respectively to highlight their relative onset based on different $\eta$.}
\label{fig:hyperbola} 
\end{figure}

To understand the significance of this representation, we see the behavior of a simple linear damped system with constant damping factor $\mathcal{D}$ and constant forcing factor $\mathcal{F}$. The differential equation can be written as: $\dot{\bar{h}}=-\mathcal{D}\bar{h}+\mathcal{F}$. We solve for this system for different values of $\mathcal{D}$ and $\mathcal{F}$, as shown in Fig.~\ref{fig:hyperbola}, which illustrates the behavior of a damped system that ``rises'' faster and begins damping earlier to reach the equilibrium for larger values of the parameter $\eta = \mathcal{D}\cdot \mathcal{F}$. Such universal behavior of a linearly damped system hints at the importance of $D$ and $F$ in governing the solution of Eq.~(\ref{eq:damped}) for capillary rise for different geometries. The general solution to the equation $\dot{\bar{h}}=-\mathcal{D}\bar{h}+\mathcal{F}$ can be written as $\bar{h}(t) = \frac{\mathcal{F}}{\mathcal{D}}(1 - e^{-\mathcal{D}t}) + h_0e^{-\mathcal{D}t}$, where $h_0$ represents the initial condition at $t=0$. The steady-state solution is obtained by taking the limit as: $\bar{h}_{eq} = \lim_{t \to \infty} \bar{h}(t) = \frac{\mathcal{F}}{\mathcal{D}}$. This equilibrium height represents the balance between forcing and damping effects. Furthermore, the initial rise rate can be computed by evaluating the time derivative as: $\left.\frac{dh}{dt}\right|_{t=0} = \mathcal{F} - \mathcal{D}h_0$. For $h_0 \ll \mathcal{F}/\mathcal{D}$, the initial dynamics are dominated by the forcing term $\mathcal{F}$ where the initial acceleration of the system is $\propto \eta$. For higher values of $\eta$, the time to attain equilibrium is also reduced while achieving a faster initial rise. Such a system is also present in electrical circuits where $\eta$ defines the circuit's dynamic response and determines how quickly the current can change, which affects the power distribution between components. The relative values determine the initial inductive losses along with the increased resistive dissipation increases as current builds. This demonstrates that despite the apparently opposing nature of damping and forcing effects, their product emerges as a natural parameter controlling system behavior. 

\begin{figure*}[t] 
    \centering
    \includegraphics[width=\textwidth]{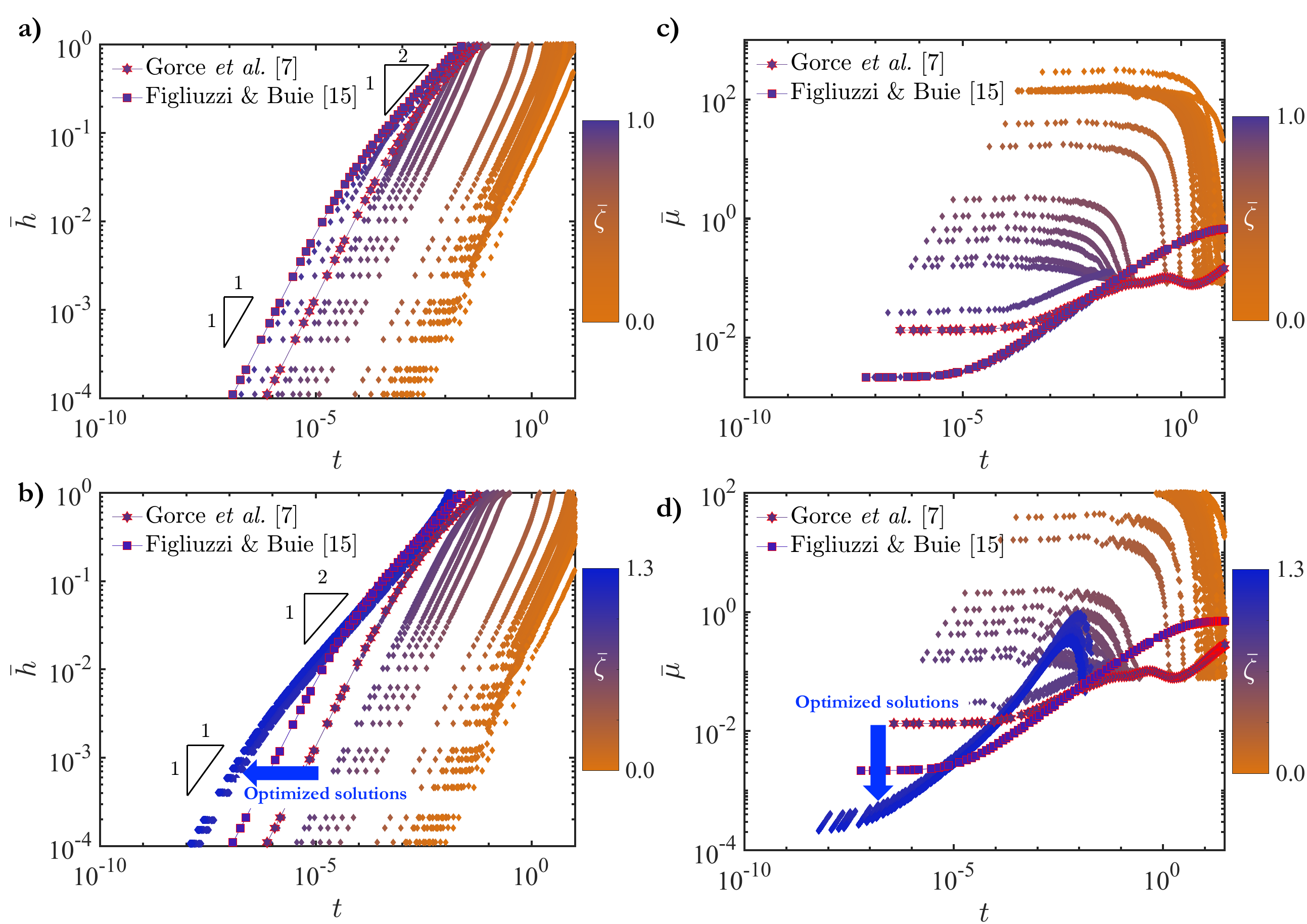} 
    \caption{Non-dimensional height ((a) and (b)) and viscosity ((c) and (d)) as function of time for capillaries with varying shapes and contact angles. Colors represent the parameter \(\bar{\zeta}\), which combines damping and forcing contributions arising from channel properties, as defined by Eq.~(\ref{eq:DF}). Capillaries designed for flow enhancement, as proposed by \citet{baptiste} and \citet{figliuzzi2013rise}, are evaluated using our framework for comparison. Capillaries optimized for flow enhancement using Eq.~(\ref{eq:ga}) are shown to enhance wicking in (b) and reduced ``effective viscosity'' in (d), beyond the state-of-the-art. These solutions are characterized by higher \(\bar{\zeta}\) values.}
    \label{fig:main}
\end{figure*}
Next, we introduce a radius profile that enables modeling of diverse channel geometries, including straight, sinusoidal, divergent, convergent, and power-law. The linearly varying component is defined as $\hat{R}(x)=R_1+(R_2-R_1)x/l$, and the amplitude of the sinusoidally varying component $\tilde{R}(x)=R_3+(R_4-R_3)x/l$, combined with the wavelength of variation $\lambda$ to obtain $R=\hat{R}+\tilde{R}\mathrm{cos}(n\pi/2 - 2\pi x/\lambda)$ where $n\in[0,1]$ introduces an additional phase shift. 

 We have calibrated the parameters $\hat{R}$, $\tilde{R}$, and $\lambda$ to fit the power-law form $R(h)=(1-h/l)^{5/6}$, matching the optimal rise predicted in \cite{baptiste}. In the system defined in Eq.~(\ref{eq:damped}), the damping and forcing coefficients exhibit a weak dependence on the height of the liquid as it rises. 
 
 Analogous to the constant damping and forcing factor analysis involving the $\eta$ parameter, we introduce a more general parameter $\zeta = \langle D \rangle  \cdot \langle F \rangle$ where the damping and forcing terms are height averaged. The governing integro-differential equation Eq.~(\ref{eq:damped}) is solved numerically using the fourth-order Runge-Kutta method for liquid height, coupled with a trapezoidal method for evaluating non-dimensional viscosity. We normalize $\zeta$ using the maximum value such that $\bar{\zeta}\in[0,1]$ and use it to interpret the solution of $\bar{h}(t)$ obtained without averaging $D$ and $F$. Our solution shows strong agreement with previous studies on purely sinusoidal channels with contact angle $0^\circ$ \cite{si}. While our formulation excludes kinetic energy contributions, this approximation is justified by the low Reynolds number regime and negligible fluid mass under the given geometric constraints. This approach aligns with standard methods in capillary rise analysis, even for varying cross-sectional areas and geometry optimization studies \cite{figliuzzi2013rise,henryk}.

    The complete solution of Eq.~(\ref{eq:damped}) reveals the temporal evolution of non-dimensionalized rise height across various geometric and flow parameters, as illustrated in Fig. \ref{fig:main}(a). The different curves correspond to channels characterized by radius variation, hydrophobic and hydrophilic surface wettability ($\theta~\in [0^\circ,104.5^\circ]$) and external pressure conditions. The different curves are observed to exhibit unique $\bar{\zeta}$ values which are used to define the colormap. The non-dimensional viscosity is also shown for different scenarios in Fig. \ref{fig:main}(c). We observe different viscosity curves for different rise dynamics with distinct $\bar{\zeta}$ as well. 
    Our analysis includes validation and comparison with previously studied channel geometries, specifically those exhibiting a power-law relationship between the radius and axial position, defined as $R(h)=(1-h/l)^{5/6}$ \cite{baptiste}, and those following a polynomial relationship \cite{figliuzzi2013rise}. These scenarios are found in the leftmost region of the curves, characterized by higher values of $\bar{\zeta}$, which aligns with the expected enhancement in wicking dynamics. These geometric configuration, characterized by a gradually decreasing radius, demonstrate notably faster rise rates compared to uniform channels. The height evolution curves exhibit behavior characteristic of linear damped systems, where $\bar{\zeta}$ serves as a key parameter governing both rise rate and equilibrium attainment. Specifically, higher values of $\bar{\zeta}$ correlate with accelerated initial rise rate and earlier equilibration which lead to overall enhanced wicking performance. Of particular interest is the rightmost region corresponding to inhibited rise. These cases typically correspond to hydrophobic surface conditions (contact angles approaching $104.5^\circ$), with a non-zero external pressure and overall higher flow resistance. This highlights the capability of $\bar{\zeta}$ to encapsulate underlying differences in flow responses through a singular value. The monotonic relationship between rise height and $\bar{\zeta}$ across the entire design space further strengthens its analogy with linear damped systems, aligning with the monotonic behavior of the parameter $\eta$ observed at varying heights prior to equilibrium, as shown in Fig.~\ref{fig:hyperbola}

\par The variation in the non-dimensional viscosity $\bar{\mu}$ across different scenarios, as shown in Fig.\ref{fig:main}(c), indicates that faster-rising flows begin with lower values of $\bar{\mu}$, which increase as viscous effects dominate near equilibrium. From Fig.\ref{fig:main}(a), we observe that faster rises transition to the viscous regime earlier, reaching equilibrium sooner and corresponding to higher $\bar{\zeta}$. Conversely, in the rightward region of Fig.\ref{fig:main}(a), lower $\bar{\zeta}$ values are associated with slower rises and higher ``effective viscosity'', indicating flow inhibition. Near equilibrium, Fig.\ref{fig:main}(c) shows a reduction in viscous effects. Among the configurations studied, the geometries proposed in \cite{baptiste,figliuzzi2013rise} exhibit the lowest $\bar{\mu}$, reflecting minimal flow resistance that serves as a key underlying response to enhance wicking.

\par The manual exploration of various geometries, external pressures, and contact angles revealed solutions with a consistent monotonic dependence on $\bar{\zeta}$. These solutions, spanning both flow inhibition and expedited rise, were compared to geometries proposed in \cite{baptiste,figliuzzi2013rise}, demonstrating that  $\bar{\zeta}$ encompasses the entire spectrum from wicking to anti-wicking. To leverage the flexibility of our parametric channel geometry, we implemented an optimization routine targeting enhanced rise rates, assuming a contact angle of $0^\circ$ and no external pressure. The monotonic dependence on $\bar{\zeta}$ underscores its importance as a key parameter distinguishing wicking from anti-wicking rise dynamics.

\par
\begin{figure}[h!]
\includegraphics[width=0.48\textwidth]{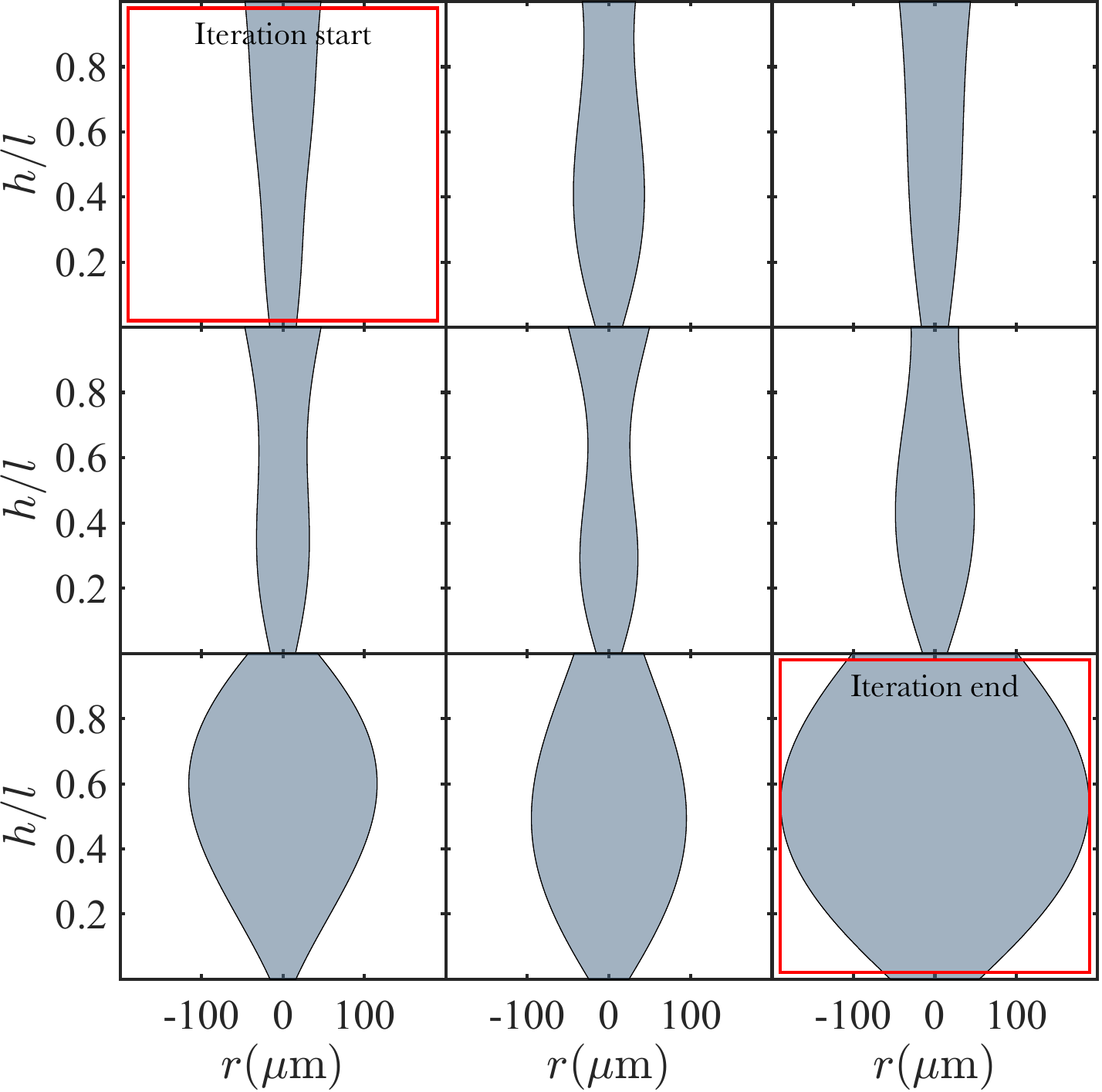} 
\caption{Evolution of the channel surface through Genetic Algorithm iterations, leading to the maximization of $\bar{\zeta}$. A truncated version of the shape evolution is shown, highlighting the initial and final shapes.}
\label{fig:shape_ga} 
\end{figure}

 A genetic algorithm-based \cite{holland1992genetic} stochastic optimizer was developed to iteratively evolve the geometry variable population based on flow solution fitness. The optimization results, aimed at maximizing $\bar{\zeta}$ as defined in Eq.(\ref{eq:ga}), are compared with existing capillary dynamics, and illustrated in Fig.\ref{fig:main}(b) and (d).
\begin{eqnarray} 
\max_{\mathbf{x}} \quad J(\mathbf{x})  = \langle D(\mathbf{x}) \rangle \cdot \langle F(\mathbf{x}) \rangle :&&
\quad \mathbf{x} = \bar{h}(\bar{R}(x), \tilde{R}(x), n, \lambda)\nonumber \\
x_i^{\mathrm{min}}  \leq x_i  \leq x_i^{\mathrm{max}}, &&\quad i = 1, \ldots,4 
\label{eq:ga}
\end{eqnarray}
The resulting channel geometry $R(x)$ from Eq.(\ref{eq:ga}), shown in Fig.~\ref{fig:shape_ga}, evolves from an initial diverging channel design and iterates to yield an optimized geometry. The shapes are evolved until convergence is achieved, minimizing the variation in the cost function over 1000 generations, with $\Delta J(\mathbf{x}) < 1e-6$.

\par The optimized height solution, $\bar{h}$, from Eq.~(\ref{eq:damped}), obtained using the genetic algorithm in Eq.~(\ref{eq:ga}), is presented in Fig.~\ref{fig:main}(b) alongside previous results. The optimized geometries demonstrate significantly enhanced liquid rise, characterized by both an earlier onset of the viscous regime and faster overall rise rates. The monotonic nature of $\bar{\zeta}$ is used to predict channels that enhance liquid rise, which is reflected in the colormap, where the new curves from the optimized geometries are placed in sequence with earlier designs. The values of $\bar{\zeta}$ for the optimized geometries exceed 1, using the same non-dimensionalization as before, indicating a $\sim 30\%$ increase. These optimized rise solutions surpass the heights predicted by \cite{baptiste,figliuzzi2013rise}. As observed in the previous analysis of faster rise cases, the onset of viscous effects occurs earlier, with an earlier transition to the $\propto \sqrt{t}$ regime from the inertial regime of $\propto t$. 

\par The analysis of non-dimensional viscosity evolution provides a quantitative measure of the underlying interplay of forces in optimized capillaries. Fig.~\ref{fig:main}(d) shows that the newly proposed geometries minimize the effective viscous resistance in the system, enabling faster liquid rise. The early onset of the viscous regime is evidenced by a rapid increase in $\bar{\mu}$, surpassing the results simulated using the geometries from \cite{baptiste,figliuzzi2013rise}. The enhanced rise is characterized by an initially low $\bar{\mu}$, which increases to a peak value, reflecting the growing viscous effects as the liquid accelerates. After gaining momentum in the initial rise phase, the liquid begins to slow down as it approaches equilibrium, causing a subsequent drop in $\bar{\mu}$.

\par The emergence of a singular parameter $\bar{\zeta}$ as a universal descriptor for capillary rise dynamics represents a significant advancement in understanding fluid transport through confined axisymmetric geometries that can differentiate rise dynamics across multiple configurations. We have demonstrated that $\bar{\zeta}$ can define the spectrum ranging between wicking and anti-wicking.
Our findings demonstrate that this phenomenon can be effectively modeled as an approximate linear damped system, where the liquid rise is governed by the interplay between forcing and damping in the system. This parameter effectively captures the complex interplay between channel geometry, surface wettability, and competing forces, providing a unified framework for characterizing diverse rise behaviors. The channel geometry, which strongly influences liquid behavior, is approximated by a linear combination of a linearly varying mean radius and a sinusoidal waveform with a varying amplitude and phase. This geometric representation encapsulates most axisymmetric shapes studied in literature and can be extended through additional truncation terms, analogous to a Fourier series expansion. The genetic algorithm developed to utilize this parameter to discover new geometries has been able to outperform previously proposed optimal shapes for wicking, thereby expanding the design space for capillary systems. Furthermore, the systematic relationship between the ``effective viscosity'' evolution and rise performance offers new insights into understanding the underlying response of the system that yields enhanced liquid transport. The wide flow response obtained through hydrophobic and hydrophilic channels could be particularly valuable in designing microfluidic systems where precise control over fluid transport rates is crucial. Furthermore, this damped system framework enables the formulation of an optimization routine, with a cost function incorporating both averaged damping and forcing effects. This approach opens new possibilities for controlling liquid flow in capillaries, offering both enhancement and tempering strategies that can work synergistically to achieve optimal control across diverse geometric configurations.\\~\\ \indent This work was supported in part by the Cornell Atkinson Center for Sustainability. We acknowledge valuable discussions with Joshua Krsek and Giancarlo D'Orazio.

\bibliography{bibfile}
\bibliographystyle{apsrev4-1}

\end{document}